\documentclass[final,1p,times]{elsarticle}
\usepackage{graphicx,amsmath,amsthm,amssymb,amsfonts}
\usepackage[bookmarks=false,pdfstartview=FitH,hyperindex=true, colorlinks, linkcolor=black, citecolor=black]{hyperref}
\journal{Nuclear Physics B}

\newcommand{\calA}{ {\mathcal A}}
\newcommand{\calC}{ {\mathcal C}}

\newcommand{\calN}{ {\mathcal N}}

\newcommand{\calR}{ {\mathcal R}}
\newcommand{\calS}{ {\mathcal S}}

\newcommand{\Rcan}{R^{\rm can}}
\newcommand{\ycan}{y_{\rm can}}

\newcommand{\hR}{\hat{R}}
\newcommand{\hRcan}{\hat{R}^{\rm can}}
\newcommand{\Lmin}{L_{\rm min}}

\begin{document}

\begin{frontmatter}
\title{Universal Critical Wrapping Probabilities \\ in the Canonical Ensemble}

\author[ustc,itp]{Hao Hu}
\ead{huhao@mail.ustc.edu.cn}
\author[ustc,itp]{Youjin Deng\corref{cor1}}
\ead{yjdeng@ustc.edu.cn}
\cortext[cor1]{Corresponding author}
\address[ustc]{National Laboratory for Physical Sciences at Microscale and Department of Modern Physics, \\
University of Science and Technology of China, Hefei, Anhui 230026, China} 
\address[itp]{State Key Laboratory of Theoretical Physics, Institute of Theoretical Physics, \\ 
Chinese Academy of Sciences, Beijing 100190, China}

\begin{abstract}
Universal dimensionless quantities, such as Binder ratios and wrapping
probabilities, play an important role in the study of critical phenomena. 
We study the finite-size scaling behavior of the wrapping probability for the Potts model 
in the random-cluster representation, under the constraint that the total number of 
occupied bonds is fixed, so that the canonical ensemble applies. 
We derive that, in the limit $L \rightarrow \infty$, the critical values of the wrapping probability are different from 
those of the unconstrained model, i.e. the model in the grand-canonical ensemble, 
but still universal, for systems with $2y_t - d > 0$ where $y_t = 1/\nu$ is 
the thermal renormalization exponent and $d$ is the spatial dimension. 
Similar modifications apply to other dimensionless quantities, such as Binder ratios. 
For systems with $2y_t-d \le 0$, these quantities share same critical universal values in the two ensembles.
It is also derived that new finite-size corrections are induced.
These findings apply more generally to systems in the canonical ensemble, e.g. the dilute Potts model
with a fixed total number of vacancies. Finally, we formulate an efficient cluster-type algorithm for the canonical ensemble,
and confirm these predictions by extensive simulations.

\end{abstract}

\begin{keyword}
Universal dimensionless quantity; Canonical ensemble; Finite-size scaling; Potts model 
\PACS{05.50.+q, 64.60.an, 64.60.F-, 75.10.Hk}
\end{keyword}

\end{frontmatter}

\section{Introduction}
\label{sec-int}
The fact that critical systems tend to display asymptotic scale invariance relates
to the existence of nontrivial dimensionless quantities that are independent of
the system size, apart from finite-size corrections. 
The wrapping probability is an example. It is defined in graphical representations 
as the probability that there exists a connected component which wraps around the periodic boundaries, 
reflecting topological properties of the system under study.
Various wrapping probabilities have been studied for 
the percolation model \cite{LPPA, Pinson94, ZLK99, NZ, WZZGD, FDB, HBD}, 
the Potts model \cite{Arguin02, MS09, TB14}, loop models \cite{LDGB} etc.
They can be related to emergent states of matter such as superfluidity 
-- the superfluid density can be calculated using the winding number \cite{PC87}. 
Many dimensionless quantities are universal at the critical point, 
in the sense that their critical values are independent of the short-range interactions, lattice types etc. 
Though their values still depend on factors such as the system shape and boundary conditions, 
the universal dimensionless quantities can be used to 
characterize universality classes of continuous phase transitions \cite{PHA, PV02}. 
The universality of the wrapping probability has been proven to depend 
only on the aspect ratio and the boundary twist of the system \cite{LPPA, Pinson94, ZLK99}. 
Wrapping probabilities are also very useful in determining the critical temperature 
for continuous phase transitions \cite{NZ, WZZGD}. 
The well-known Binder ratio \cite{Binder81} is another kind of universal dimensionless 
quantities whose properties have been widely studied \cite{BinderStu}.

In a recent work \cite{HBD}, Hu et al.\ studied the bond (site) percolation model with 
a fixed total number of occupied bonds (sites). The system is referred to 
as in the canonical ensemble (CE), while the model without the constraint is said
to be in the grand-canonical ensemble (GCE).
It is found that,  while the wrapping probabilities and dimensionless ratios share the same values 
within each of the two different ensembles, some universal quantities in the GCE become nonuniversal in the CE, 
such as the excess cluster number. 
The percolation model is special, since the occupation of each edge is independent of other edges
and the average bond density has no finite-size dependence. 
Thus we ask whether the universality of observables depends on the ensembles in other models, 
for which there exist thermal fluctuations that are accompanied by finite-size dependence of the particle density.  
We choose to study the Potts model \cite{Wu82} in the random-cluster representation, 
in which the bonds are treated as particles.
Our emphases are on the universality and finite-size scaling (FSS) of the wrapping probability in the CE, 
since their universal values at the critical point are exactly known in the GCE \cite{Arguin02}, 
and their FSS in the GCE is well understood. 
Moreover, the fact that the wrapping can be defined for each configuration significantly
simplifies theoretical derivations, as shown later in the text.

For systems in the CE, there exists a constraint that the total number of particles is fixed. 
From Fisher renormalization theory \cite{Fisher68}, under this kind of constraint, 
the system changes its way approaching the critical point as the temperature goes to 
its critical value, comparing with the asymptotic behavior in the GCE. 
There is a relation between the temperature field in the CE 
and that in the GCE. It turns out that there may exist universal changes for critical exponents 
in such constrained systems: 
if the specific heat is divergent in the GCE, namely $C \propto |T-T_c|^{-\alpha}$ with $\alpha>0$, 
it will become finite in the CE, with $\alpha_{\rm can} = - \alpha/(1-\alpha)$, 
where ``{\rm can}" indicates quantities in the CE. 
Other critical exponents may also be renormalized when $\alpha>0$, e.g. $\beta_{\rm can} = \beta/(1-\alpha)$, 
$\gamma_{\rm can} = \gamma/(1-\alpha)$, $\nu_{\rm can} = \nu/(1-\alpha)$, where $\beta$, $\gamma$ and $\nu$
are exponents for the magnetization, susceptibility and correlation length, respectively. 
Fisher renormalization has been extensively studied and applied to explain experimental results for
constrained systems (see e.g. \cite{HBD, PV02, IK, KHF, DengThesis, Krech, MOF, MrFo} and references therein). 
However, the study of the universality of dimensionless observables in the CE 
appears to be largely neglected in the literature \cite{IK}. 

We conduct a FSS analysis for the Potts model, and support our analysis by extensive numerical simulations. 
Efficient cluster Monte Carlo algorithms are designed for the simulations in the CE .
The article is organized as follows. In Sec.~\ref{sec-mod}, we introduce the model and observables.
We apply Fisher renormalization to finite-size systems in Sec.~\ref{sec-fr}. 
The universal change of the wrapping probability, and new finite-size corrections are derived in Sec.~\ref{sec-uc}.
Sec.~\ref{sec-num} presents simulation results that verify our derivations.
A brief summary and discussion is given in Sec.~\ref{sec-sum}.

\section{Model and Observables}
\label{sec-mod}
We introduce the Potts model \cite{Wu82} and define the observables in this section.

For a lattice $G$ with edge set $\{ e_{ij} \}$, 
the $q$-state Potts model is defined by the Hamiltonian
\begin{equation}
{\cal H }/k_B T = -K \sum_{e_{ij}}  \delta_{\sigma_i \sigma_j} \, , 
\end{equation}
where $\sigma_i$ takes one of the colors $1, 2, \cdots,q$, representing a spin at site $i$. 
Ferromagnetic couplings $K>0$ occur between nearest-neighbor Potts spins. 
Performing the Kasteleyn-Fortuin (KF) mapping \cite{KF},
which puts a bond on each edge $e_{ij}$ with probability $p=(1-e^{-K}) \delta_{\sigma_i \sigma_j}$,
one gets the partition sum of the Potts model in the random-cluster (RC) representation as
\begin{equation}
Z_{\rm RC}(v,q) = 
\sum_{\calA \subseteq G} v^{\calN_{\rm b}} q^{\calN _{\rm c}}
\;\;\; (v= e^K-1),
\label{zrc}
\end{equation}
where the summation is over all subgraphs $\calA$ of $G$. 
A cluster on a subgraph is defined as a connected component consisting of sites and bonds.
$\calN _{\rm b}$ and $\calN _{\rm c}$ are the number of occupied bonds and clusters, respectively. 
The parameter $q$ can take real positive values in the RC representation. 
In two dimensions, the model undergoes a continuous phase transition for $q \le 4$. 
The critical point on the square lattice is predicted to be $v_c=\sqrt{q}$,
with the critical bond density $\rho_c = 1/2$. 
The occupation of different edges are correlated generally for $q \ne 1$. 
In the limit $q \rightarrow 1$, the Potts model reduces to the bond percolation model,
for which the presence of bonds at different edges are independent. 

In the RC representation, we study the wrapping probabilities $R_x$, $R_1$, $R_b$, $R_e$: 
$R_x$ counts the probability that there exists a cluster connecting to itself along
the $x$ direction, irrespective of the $y$ direction; $R_1$ is for along the $x$ direction,
and with no cluster connecting to itself along the $y$ direction; $R_b$ is for simultaneous
wrapping in both directions; and $R_e$ is for wrapping in at least one direction.
Since $R_e=2R_1+R_b$ and $R_x=R_1+R_b$, only two of these probabilities are
independent. Exact results for these wrapping probabilities can be obtained from 
Ref.\cite{Arguin02} for Potts model in the GCE. 

The relation between wrapping probabilities in the CE and the GCE is presented as follows.
Denoting $\calR=1 \; (0)$ for a subgraph with (without) a wrapping cluster,
we define the average of a wrapping probability $R^{\rm can}$ for a finite
lattice with linear extension $L$ in the CE as
\begin{equation}
R^{\rm can} (\rho, L) = 
\frac{
\sum_{\calA_\rho} \calR q^{\calN_{\rm c}}
}{
\sum_{\calA_\rho} q^{\calN_{\rm c}}
} \; , 
\label{RCE}
\end{equation}
where the summation is on all subgraphs $\calA_\rho$ which have 
$\calN_{\rm b} = \rho N_e$ bonds. Here, $\rho$ is the bond density, 
and $N_e$ the total number of edges. 
The wrapping probability in the GCE is defined as
\begin{equation}
R (v, L) = 
\frac{
\sum_{\rho=0}^{1} 
v^{\calN_{\rm b}} \sum_{\calA_\rho} \calR q^{\calN_{\rm c}}
}{
\sum_{\rho=0}^{1} 
v^{\calN_{\rm b}} \sum_{\calA_\rho} q^{\calN_{\rm c}}
} \; . 
\label{RGCE}
\end{equation}
With the definition
\begin{equation}
F(\rho; v, L) =
\frac{ 
v^{\calN_{\rm b}} \sum_{\calA_\rho} q^{\calN_{\rm c}}
}{
\sum_{\rho=0}^{1} 
v^{\calN_{\rm b}} \sum_{\calA_\rho} q^{\calN_{\rm c}}
} \; ,
\end{equation}
it can be obtained from Eqs.~(\ref{RCE}) and (\ref{RGCE}) that 
\begin{equation}
R(v, L)=\int {\rm d} \rho R^{\rm can}(\rho,L) F(\rho; v, L) \;, 
\label{RRcan}
\end{equation}
where the integral denotes summation over all possible values of $\rho$.
The quantity $F(\rho; v, L) d \rho$ is the grand-canonical probability that,
for given parameters $(v,L)$, a configuration has a bond density $\rho$. 

\section{Fisher renormalization in finite-size systems}
\label{sec-fr}

\subsection{Finite-size scaling in the grand-canonical ensemble}
\label{sec-fss-gce}
For the Potts model in the GCE, 
it is expected \cite{PF84} that near the critical point $v_c$ the grand potential scales as 
\begin{equation}
\Omega(v, L) \simeq \Omega_r(v) + L^{-d} \Omega_s(tL^{y_t}),
\end{equation}
where $\Omega_r$ and $\Omega_s$ are the regular and singular parts of the potential, respectively. 
The function $\Omega_s$ is universal. 
The parameter $t$ is the thermal scaling field, and $y_t = 1/\nu$ is the associated exponent.
One has $t \simeq A (v-v_c)$, with $A$ being a nonuniversal metric factor. 
The magnetic scaling field and irrelevant scaling fields are omitted for simplicity.

The average bond density is 
\begin{eqnarray}
\bar{\rho}(v,L) = - \frac{\partial \Omega(v,L)}{\partial v} 
\simeq \rho_r(v) + AL^{y_t-d} \rho_s(tL^{y_t}) \;,
\label{grand-rho}
\end{eqnarray}
where $\rho_s(tL^{y_t})=-\Omega_s'(tL^{y_t})$, $\rho_r(v)=-\Omega'_r(v)$.
And its fluctuation scales as
\begin{equation}
\langle (\rho - \bar{\rho})^2 \rangle 
\simeq L^{-d} ( \rho_r'(v) + A^2 L^{2y_t-d} \rho_s'(tL^{y_t}) ) \;.
\label{eq-rho2}
\end{equation}

The wrapping probability scales as
\begin{equation}
R(v, L) \simeq \hat{R}(tL^{y_t}) \;,
\label{hatR-gce}
\end{equation}
where $\hat{R}$ is a universal function.

With $\rho_c \equiv \bar{\rho}(v_c,L \rightarrow \infty) = \rho_r (v_c)$,
near $v=v_c$, one can define 
\begin{eqnarray}
\Delta \rho = \bar{\rho}(v, L) - \rho_c  
\simeq (\rho_r' + A^2L^{2y_t-d}\rho_s') \Delta v + AL^{y_t-d}\rho_s \;, 
\label{drho}
\end{eqnarray}
where $\Delta v = v-v_c$, $\rho_r'=\rho_r'(v_c)$, $\rho_s'=\rho_s'(0)$ and $\rho_s=\rho_s(0)$. 
This is obtained by Taylor expanding Eq. (8) around $v_c$ and ignoring non-linear terms.

Substituting Eq.~(\ref{drho}) into $t \simeq A \Delta v$, 
one derives that
\begin{eqnarray}
t \simeq \frac{A \Delta \rho - A^2 L^{y_t-d} \rho_s}{\rho_r'+A^2L^{2y_t-d}\rho_s'} \;.
\label{t-tprime}
\end{eqnarray}

In the two-dimensional $q=2$ (Ising) case one has $2y_t-d=0$, and the system in the GCE has a logarithmic specific-heat
anomaly which scales as $C_{L=\infty}(t) \sim |\ln {|t|}|$ \cite{Onsager44}.
At the critical point, Eq.~(\ref{eq-rho2}) is replaced by 
\begin{equation}
\langle (\rho - \bar{\rho})^2 \rangle 
\simeq L^{-d} ( C_0 + C_1 \ln L ) \;,
\label{eq-rho2-log}
\end{equation}
where $C_0$, $C_1$ are nonuniversal constants.
We note that, from Eq.~(\ref{eq-rho2}) to Eq.~(\ref{eq-rho2-log}),
$C_1$ contains a contribution from $\rho_s'$, thus the logarithmic factor $\ln L$ 
is associated with $\rho_s'=-\Omega_s''(0)$, i.e. the second derivative of the singular part of the free energy. 
Accordingly, Eq.~(\ref{drho}) changes to 
\begin{equation}
\Delta \rho \simeq (C_0 + C_1 \ln L ) \Delta v + AL^{y_t-d}\rho_s \;. 
\label{drhoLog}
\end{equation}
It follows that
\begin{eqnarray}
t \simeq \frac{A \Delta \rho - A^2 L^{y_t-d} \rho_s}{ C_0 + C_1 \ln L } \;.
\label{t-tprime-Log}
\end{eqnarray}

\subsection{Finite-size scaling in the canonical ensemble}
\label{sec-fss-ce}

In the CE, we {\it conjecture} that near the critical density $\rho_c$, 
the Helmholtz free energy scales as
\begin{eqnarray}
f(\rho,L) \simeq f_r(\rho)+L^{-d}f_s(x(\rho, L)) \;.
\end{eqnarray}
In principle, $f$ is related to the grand potential by a Legendre transform $f = \Omega + v \rho$,
with $\rho=\bar{\rho}(v,L)$.
Thus we {\it suppose} that in the CE, $x(\rho, L)$ is given by $t L^{y_t}$.

For $2y_t-d>0$, with $\rho_r' + A^2 L^{2y_t-d} \rho_s' \simeq A^2 L^{2y_t-d}\rho_s'$ as $L$ becomes very large, it can be derived from Eq. (\ref{t-tprime}) that, 
\begin{eqnarray}
x(\rho, L) = t L^{y_t} \simeq \Delta \rho L^{d-y_t}/A\rho_s' - \rho_s/\rho_s'  
= \tau L^{y_\tau} - \delta \;,
\label{tauYtau}
\end{eqnarray}
where $\tau = B \Delta \rho$ with $B=1/A\rho_s'$, $y_\tau = d-y_t$, and $\delta =  \rho_s/\rho_s'$.
The parameter $\delta$ is universal, since both $\rho_s$ and $\rho_s'$ are universal quantities.
From Eq. (\ref{tauYtau}), as $L \rightarrow \infty$, $\tau=0$ corresponds to $t=0$, 
but $t L^{y_t}=0$ leads to  $\tau L^{y_\tau}=\delta$. 
We note that the hyperscaling relations $\alpha = 2-d/y_t$ and $\alpha_{\rm can} = 2-d/y_\tau$,
when combined with $y_\tau = d-y_t$, reproduce the 
Fisher renormalization result $\alpha_{\rm can} = -\alpha / (1-\alpha)$.

Thus, the wrapping probability should scale as
\begin{equation}
\Rcan(\rho, L) \simeq \hRcan (\tau L^{y_\tau} - \delta) \;,
\label{Rcan}
\end{equation}
which at $\rho = \bar{\rho}(v,L)$ leads to 
\begin{eqnarray}
\partial^2 R^{\rm can} (\rho, L) / \partial \rho^2 
\simeq B^2 L^{2y_\tau} \hR^{\rm can ''}(\tau L^{y_\tau} - \delta) 
= L^{2(d-y_t)} \hR^{\rm can ''}(\tau L^{y_\tau} - \delta)/A^2 \rho_s'^2 .
\label{eq-Rcan2}
\end{eqnarray}

For $2y_t-d<0$, from Eq. (\ref{t-tprime}), one gets
\begin{eqnarray}
x(\rho, L) = t L^{y_t} \simeq A \Delta \rho L^{y_t}/\rho_r' 
= \tau L^{y_\tau} ,
\end{eqnarray}
where $\tau = B \Delta \rho$ with $B = A / \rho_r'$, $y_\tau = y_t$.
The wrapping probability takes the same form as Eq. (\ref{Rcan}) with $\delta = 0$, 
and Eq. (\ref{eq-Rcan2}) is replaced by 
\begin{eqnarray}
\partial^2 R^{\rm can} (\rho, L) / \partial \rho^2 
\simeq B^2 L^{2y_\tau} \hR^{\rm can ''}(\tau L^{y_\tau})
= L^{2 y_t} \hR^{\rm can ''}(\tau L^{y_\tau}) A^2 / \rho_r'^2 .
\label{eq-Rcan2-AlphaN}
\end{eqnarray}
 
For $2y_t-d=0$, from Eq. (\ref{t-tprime-Log}), one has 
\begin{eqnarray}
x(\rho, L) = tL^{y_t} \simeq A \Delta \rho L^{y_t} / C_1 \ln L 
= \tau L^{y_\tau} / \ln L \;,
\label{tauYtauLog}
\end{eqnarray}
where $\tau = B \Delta \rho$ with $B=A/C_1$, $y_\tau = y_t$. 
Thus the singular part of the free energy changes to $f_s(\tau L^{y_\tau} / \ln L)$,
and the wrapping probability scales as 
\begin{equation}
\Rcan(\rho, L) \simeq \hRcan (\tau L^{y_\tau} / \ln L) \;,
\end{equation}
leading to
\begin{eqnarray}
\partial^2 R^{\rm can} (\rho, L) / \partial \rho^2
\simeq B^2 L^{2y_\tau} \hR^{\rm can ''}(\tau L^{y_\tau}) / (\ln L)^2 
= L^{2 y_t} (\ln L)^{-2} \hR^{\rm can ''}(\tau L^{y_\tau} / \ln L) A^2 / C_1^2 \;.
\label{eq-Rcan2-AlphaN}
\end{eqnarray}

We also note that, close to $\rho_c$, as $L \rightarrow \infty$, 
one has $|\ln |\Delta \rho|| \propto \ln L$.
Thus with $\tau = B \Delta \rho$, Eq.~(\ref{tauYtauLog}) leads to
\begin{eqnarray}
|t| \propto |\tau|  |\ln |\tau| |^{-1},
\end{eqnarray}
which is the Fisher renormalization result
in the thermodynamic limit \cite{Fisher68, KHF} for the logarithmic case.

\section{Universal difference and finite-size corrections}   
\label{sec-uc}

Expanding the right-hand side of Eq.~(\ref{RRcan}) near 
$\rho=\bar{\rho}(v, L)$, one gets
\begin{eqnarray}
R(v, L)
\simeq R^{\rm can}(\bar{\rho}, L)  
+ \frac{\partial^2 R^{\rm can}(\bar{\rho}, L)}{2 \partial \rho^2} \langle(\rho- \bar{\rho})^2\rangle ,
\label{expand-R}
\end{eqnarray}
where the first order term vanishes since $\langle \rho - \bar{\rho} \rangle = 0$.
We shall make use of this equation to study the difference between critical wrapping probabilities 
in the CE and the GCE, which are defined as 
$R^{\rm can}_0 \equiv R^{\rm can} (\rho_c, L \rightarrow \infty)$
and 
$R_0 \equiv R(v_c, L \rightarrow \infty$), respectively.

\subsection{$2y_t-d>0$}
For finite-size systems at the critical point $v_c$, 
one has $t L^{y_t}=0$, and from Eq. (\ref{tauYtau}) $\tau L^{y_\tau} \simeq \delta$.
From Eqs.~(\ref{hatR-gce}) and (\ref{Rcan}), one obtains $R(v_c, L) \simeq \hR(0)$
and $\Rcan(\rho_c, L) \simeq \hRcan(-\delta)$,
where the latter value is different from $\Rcan(\bar{\rho}(v_c, L), L) \simeq \hRcan(0)$. 
In the limit $L \rightarrow \infty$, these approximate relations become exact, 
thus one gets $R_0 =  \hR(0)$  and $\Rcan_0 = \hRcan(-\delta)$.

The combination of Eqs. (\ref{eq-rho2}) and (\ref{eq-Rcan2}) at the critical point $v_c$ 
leads to  
\begin{eqnarray}
\frac{\partial^2 R^{\rm can}(\bar{\rho}, L)}{2 \partial \rho^2} \langle(\rho- \bar{\rho})^2\rangle  
=  \frac{\hR^{\rm can ''}(0) }{2 \rho_s'} 
+ \frac{L^{d-2y_t} \hR^{\rm can ''}(0) \rho_r'}{2 A^2 \rho_s'^2} \;.
\end{eqnarray}
Substituting the above results into Eq. (\ref{expand-R}), we derive 
\begin{eqnarray}
R_0  \simeq \hR^{\rm can}(0) +  
\frac{\hR^{\rm can ''}(0) }{2 \rho_s'} 
+ \frac{L^{d-2y_t} \hR^{\rm can ''}(0) \rho_r'}{2 A^2 \rho_s'^2} \;.
\label{RRcan0}
\end{eqnarray}

Equation~(\ref{RRcan0}) tells that: 
(i) the critical values of the wrapping probability in the GCE are different from the values in the CE. 
This remains true in the limit of $L \rightarrow \infty$,   
i.e. $R_0 \neq \Rcan_0 = \hRcan(-\delta)$. And, since $R_0$,
$\rho_s'$ and $\delta$ are universal, (ii) $\hRcan (0)$ and $\hR^{\rm can ''}(0)$ are universal,
as well as the difference between $\Rcan_0$ and $R_0$.
It can be derived that higher-order terms omitted in Eq. (\ref{expand-R}) lead to 
universal terms similar to $\hR^{\rm can''}(0)/2\rho_s'$, but with high-order derivatives of $\hRcan$.  
These terms also contribute to the difference between $R_0$ and $\Rcan_0$. 

The leading finite-size correction in the GCE has an exponent $y_i$,
i.e. a term proportional to $L^{y_i}$ is present on the left-hand side of Eq. (\ref{RRcan0}).
Since one usually has $y_i \ne d-2y_t$, there should exist a
correction term  $\sim L^{\ycan}$ in the CE with $\ycan=d-2y_t$, 
which cancels the term proportional to $L^{d-2y_t}$ in Eq. (\ref{RRcan0}). 

\subsection{$2y_t-d<0$}

For $2y_t-d<0$, at the critical point $v_c$, one has $\tau L^{y_\tau}=0$. 
Following a procedure similar to that for $2y_t-d>0$, we get 
\begin{eqnarray}
R_0 \simeq \Rcan _0 + \frac{\hR^{\rm can''}(0) A^2}{2 \rho_r'} L^{2y_t-d} \;.
\end{eqnarray}
Thus, at criticality we have $R_0=R^{\rm can}_0$, the wrapping probability
is identical in the two ensembles,
and a finite-size correction with exponent $\ycan = 2y_t-d$ occurs
in the CE. 

\subsection{$2y_t-d=0$}

For $2y_t-d=0$, one has $\tau L^{y_\tau} / \ln L = 0$ at the critical point $v_c$. 
It can be derived that
\begin{eqnarray}
R_0 \simeq \Rcan_0 + \frac{\hR^{\rm can''}(0) A^2}{2 C_1} / \ln L \;,
\end{eqnarray}
which tells that $R_0=\Rcan_0$, 
and a logarithmic correction term proportional to $1/\ln L$ is present in the CE.

\subsection{Remark}

The universal difference between the wrapping probability in the CE and GCE
for systems with $2y_t-d>0$ is the main finding of this work. 
It is important to note that, for this universal difference,
there is a contribution coming from fluctuations of the bond density in the GCE, 
even when the universal parameter $\delta$ is zero. 
The suppression of bond-density fluctuations also leads to finite-size corrections
with an exponent $\ycan=-|2y_t-d|$ for systems with $2y_t-d \ne 0$, 
and logarithmic corrections such as a term $\sim 1/\ln L$ for $2y_t-d=0$. 
Finite-size corrections can also come from Fisher renormalization \cite{Krech,MOF,MrFo}: 
for $2y_t-d>0$, from Eq. (\ref{t-tprime}) to (\ref{tauYtau}), 
a correction term with $L^{d-2y_t}$ should be present when higher-order effects are considered; 
similarly one has a correction term $\sim L^{2y_t-d}$ for $2y_t-d<0$, 
and $\sim 1/\ln L$ for $2y_t-d=0$.

We mention that the above findings should apply to other universal observables like 
Binder ratio and to other canonical-ensemble systems like the dilute Potts model
with a fixed total number of vacancies.

\section{Numerical verification}  
\label{sec-num}
To verify the analysis in the previous section, we conducted extensive 
Monte Carlo calculations, for which the simulation details and results are 
presented below.

\subsection{Monte Carlo  Algorithms}
For the simulation of the Potts model in the GCE,
we employ the Swendsen-Wang (SW) algorithm \cite{SW}.
In the CE, the total number of bonds is fixed and we designed 
the following algorithm for updating the configurations: 
(i) randomly distribute the bonds on the lattice edges; 
(ii) for each cluster, color it to be `k' with probability $p_k=1/q$ 
for each color ($k=1,2,...,q$); 
(iii) independently on each subgraph $G[V_k]$ ($V_k$ represents all sites with color `k', 
and $G[V_k]$ consists of $V_k$ and the edges between sites with color `k'), 
employ Kawasaki dynamics \cite{Kawasaki} for bond percolation, 
i.e. exchange the state of two randomly selected edges;
(iv) erase the colors; (v) repeat steps (ii), (iii) and (iv) until the required number of
samples is reached.
The canonical algorithm can be easily adapted to simulate RC models with non-integer
$q$ value $(q >1)$: for each cluster, color it to be `1' with probability $p=1/q$,
and `2' with $p=1-1/q$; then on subgraph $G[V_1]$, employ Kawasaki dynamics for bond percolation.
For non-integer $q$-state RC model in the GCE, 
the Swendsen-Wang-Chayes-Machta algorithm (SWCM) \cite{CM} algorithm can be used. 

\subsection{Simulation details}
The simulations were conducted on $L \times L$ square lattices with periodic boundary conditions. 
We simulated the $q=3$ Potts model which has $2y_t-d=2/5>0$, 
and the $q=2$ Potts model which has $2y_t-d=0$.
For the models in the GCE, the critical point is $v_c = \sqrt{q}$,
and the critical bond density is $\rho^b_c = \langle \calN_{\rm b} \rangle / 2L^2 = 1/2$.
Further details of the simulations at the critical point are summarized in Table \ref{tab-sim}. 
With $L=64, 128, 256$, we also did simulations for the models at several points near the critical point, 
for which over $10^7$ samples were taken at each point for each size. 

\begin{table}
\caption{Details of the simulations for the Potts models at the critical point. 
The number of updates between two samples were chosen so that the samples are approximately independent. 
For each size, the total number of samples were taken from over $50$ independent jobs, 
for which each job had lots of updates (roughly equivalent to $1/5$ of the number of samples) 
thrown for thermalization before sampling.}
\label{tab-sim}
\begin{center}
    \begin{tabular}{l|c|c}
    \hline
    Model      & $L$ 		& Number of samples   \\
    \hline
    $q=3$ GCE  & $\{8, 16, 32,64,128, 256, 512 \}$ & $3.0 \times 10^8$ \\
    $q=3$ CE   & $\{8, 16, 32,64,128 \}$ & $1.2 \times 10^9$ \\
 	       & $\{ 256 \}$ & $2.6 \times 10^8$ \\
 	       & $\{ 512 \}$ & $1.4 \times 10^8$ \\
 	       & $\{ 1024 \}$ & $4.0 \times 10^7$ \\
 	       & $\{ 2048 \}$ & $2.3 \times 10^7$ \\
    $q=2$ GCE  & $\{8, 16, 32,64,128, 256 \}$ & $5.0 \times 10^7$ \\
 	       & $\{ 512 \}$ & $2.5 \times 10^7$ \\
    $q=2$ CE   & $\{8, 16, 32,64,128, 256 \}$ & $5.0 \times 10^7$ \\
 	       & $\{ 512 \}$ & $2.5 \times 10^7$ \\
    \hline
    \end{tabular}
\end{center}
\end{table}

\subsection{Numerical results for $2y_t-d>0$}
\begin{figure}
\begin{center}
\includegraphics[width=8.0cm]{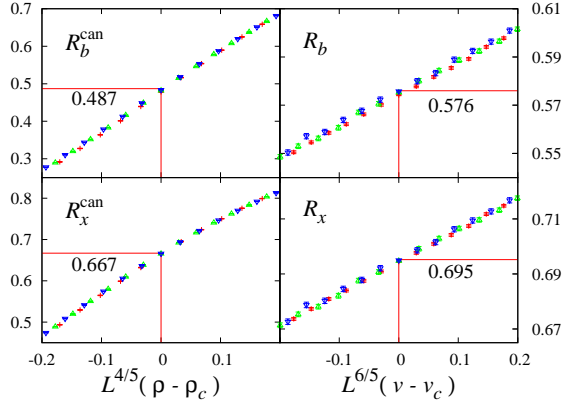}
\end{center}
\caption{Wrapping probabilities $R_b$ (top), $R_x$ (bottom) versus $L^{d-y_t}(\rho-\rho_c)$ 
in the canonical ensemble (left), and versus $L^{y_t}(v-v_c)$ in the grand-canonical ensemble (right), 
for the $q=3$ Potts model on square lattices, with $y_t=6/5$ and the spatial dimension $d=2$. 
Data points for three different sizes $L=64,128,256$ (cross, up-triangle, down-triangle) collapse well 
into a single curve. Solid lines indicate the coordinates of the critical values.}
\label{fig-RxDRho}
\end{figure}

In the CE, the thermal renormalization exponent is renormalized as $y_\tau=d-y_t=4/5$ for $q=3$ Potts model. 
From the simulation data, we plotted this scaling renormalization for two independent wrapping 
probabilities $R_x$ and $R_b$, as shown in Fig.~\ref{fig-RxDRho}.
The figure also shows that the critical wrapping probabilities in the CE are different from
those in the GCE.

Using the method of least squares, we performed fits with our Monte Carlo data 
at the critical point by the formula 
\begin{equation}
R=R_0+b_1 L^{y_1} + b_2 L^{y_2},
\label{fitForB}
\end{equation}
where $R_0$ is the universal value of the wrapping probability; 
$y_1$ and $y_2$ are the leading and subleading correction exponents, both being negative; 
and $b_1$, $b_2$ are nonuniversal amplitudes. 
For the GCE, the leading correction exponent is $y_1=y_i=-4/5$ \cite{Nienhuis87, Cardy87}. 
And for the CE, the correction exponents are expected to be $y_1=\ycan=-2/5$ and $y_2=-4/5$. 
We set a lower cutoff $L \ge L_{\rm min}$ on the data and observed the change of 
the residual $\chi^2$ as $L_{\rm min}$ increases. 
Subsequent increases of $L_{\rm min}$ do not make the $\chi^2$ value 
drop vastly by more than one unit per degree of freedom.

Table \ref{tab-wrapping-q3} summarizes the results of our fits. 
The fits were made by assuming one or two correction terms, and in alternate lines
using the predicted values of $y_1$ and $y_2$. For the GCE a fit was made 
with $R_0$ fixed at its theoretical value.
As expected from the analysis in Sec. \ref{sec-uc} , the critical values of 
the wrapping probabilities in the CE are significantly different from those in the GCE.
It is interesting to note that the predicted leading correction amplitude $b_1$ is 
consistent with $b_1 =0$, while this no longer holds for the dilute q=3 Potts 
model (see the manuscript later).”

\begin{table}
\caption{Fit results for wrapping probabilities of the $q=3$ Potts model.
The abbreviations ``GCE" and ``CE" stand for the grand-canonical and canonical ensemble, respectively.
Entries ``-" indicate that the corresponding term is not included in the fit, 
and the numbers without error bars are fixed in the fits.
Exact values for $R_0$ were obtained using formulas in Ref.~\cite{Arguin02}.
Error margins are quoted as two times of the statistical errors in the fits.}
\label{tab-wrapping-q3}
\begin{center}
    \begin{tabular}{cllllllll}
    \hline
        \multicolumn{9}{c}{Fits using $R = R_0 + b_1 L^{y_1} + b_2 L^{y_2}$} \\
    \hline
& & $R_0$ & $b_1$ & $y_1$ & $b_2$ & $y_2$ & $\Lmin$ & $\chi^2/DF$ \\
    \hline
$R_x$ & GCE & $0.695\,000\,176$ & $-0.04(2)$ & $-1.07(18)$ & - & - & $16$ & $0.2$ \\
&     & $0.695\,2(3)$ & $-0.021(12)$ & $-4/5$ & - & - & $64$ & $0.1$ \\
& CE  & $0.667\,2(3)$ & $-0.05(2)$  & $-0.69(13)$  & - & -  & $64$ & $0.2$ \\
&     & $0.667\,3(5)$ & $-0.005\,(6)$  & $-2/5$  & $-0.05(2)$ & $-4/5$  & $64$ & $0.1$ \\
$R_1$ & GCE & $0.118\,666\,330$ & $\ \; \, 0.056\,(6)$ & $-0.86(3)$ & - & - & $32$ & $0.6$ \\
&     & $0.118\,68(10)$ & $\ \; \, 0.040\,(6)$ & $-4/5$ & - & - & $128$ & $0.5$ \\
& CE  & $0.180\,16(14)$ & $\ \; \, 0.152\,(14)$  & $-0.77(3)$  & - & -  & $64$ & $0.8$ \\
&     & $0.180\,1(2)$ & $\ \; \, 0.003\,(3)$  & $-2/5$  & $0.157\,(10)$ & $-4/5$  & $64$ & $0.7$ \\
$R_b$ & GCE & $0.576\,333\,845$ & $-0.115\,(18)$ & $-0.97(6)$ & - & - & $16$ & $0.4$ \\
&     & $0.576\,6(2)$ & $-0.066\,(6)$ & $-4/5$ & - & - & $32$ & $0.1$ \\
& CE  & $0.487\,0(4)$ & $-0.20(4)$  & $-0.75(5)$  & - & -  & $64$ & $0.3$ \\
&     & $0.487\,2(6)$ & $-0.008\,(8)$  & $-2/5$  & $-0.21(3)$ & $-4/5$  & $64$ & $0.2$ \\
$R_e$ & GCE & $0.813\,666\,506$ & $\ \; \, 0.034\,(9)$ & $-0.84(9)$ & - & - & $16$ & $0.7$ \\
      &     & $0.813\,8(2)$ & $\ \; \, 0.027\,(4)$ & $-4/5$ & - & - & $32$ & $0.3$ \\
& CE  & $0.847\,30(14)$ & $\ \; \, 0.103\,(10)$  & $-0.82(4)$  & - & -  & $32$ & $0.2$ \\
&     & $0.847\,4(3)$ & $-0.001\,3(28)$  & $-2/5$  & $0.102(7)$ & $-4/5$  & $32$ & $0.2$ \\
    \hline
    \end{tabular}
\end{center}
\end{table}

\subsection{Numerical results for $2y_t-d=0$}

\begin{figure}
\begin{center}
\includegraphics[width=8.0cm]{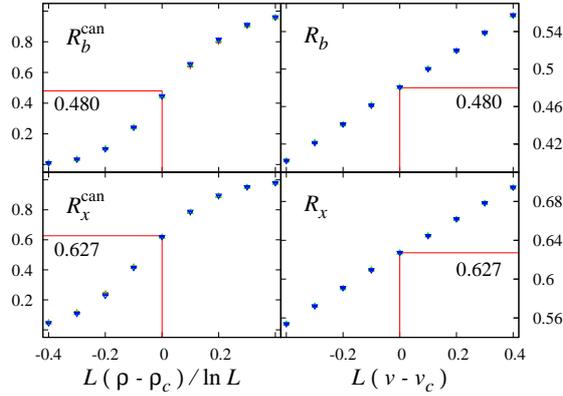}
\end{center}
\caption{Wrapping probabilities $R_b$ (top), $R_x$ (bottom) versus $L^{d-y_t}(\rho-\rho_c) / \ln L$ 
in the canonical ensemble (left), and versus $L^{y_t}(v-v_c)$ in the grand-canonical ensemble (right), 
for the Ising model on the square lattice, with $y_t=1$ and the spatial dimension $d=2$. 
Data points for three different sizes $L=64,128,256$ (cross, up-triangle, down-triangle) collapse well 
into a single curve.  Solid lines indicate the coordinates of the critical values.
At $\rho = \rho_c$, a small deviation of $R_b^{\rm can} (L)$ from its universal value $0.480$ comes from 
finite-size corrections, for which the leading term scales as $1/\ln L$.}
\label{fig-RbRxq2}
\end{figure}

For the logarithmic case $2y_t-d=0$, near the critical point, 
the analysis in Sec.~\ref{sec-fr} tells that the wrapping probabilities scale as 
$L^{y_t}(\rho - \rho_c)/\ln L$ in the CE, in contrast to $L^{y_t}(v-v_c)$ in the GCE. 
This scaling is shown in Fig.~\ref{fig-RbRxq2} for the Ising ($q=2$ Potts) model on the square lattice. 

\begin{figure}
\begin{center}
\includegraphics[width=8.0cm]{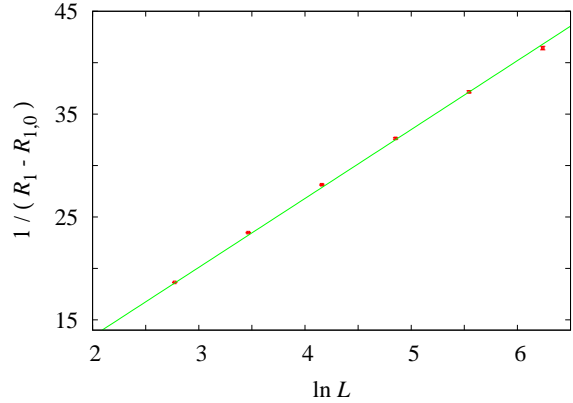}
\end{center}
\caption{Quantity $1/(R_1-R_{1,0})$ versus $\ln L$ for the Ising model 
in the canonical ensemble. The solid straight line is described by  $1/(R_1-R_{1,0}) = 6.70 \; {\ln L}$, 
with $R_{1,0}=0.146\;436\;927$ being the exact value obtained using formulas in Ref.\cite{Arguin02}. 
Small deviation of data points from the straight line can be attributed to higher-order correction terms.} 
\label{fig-R1Lq2}
\end{figure}

For the Ising model at the critical point, we performed fits to our data in the GCE with the formula
\begin{equation}
R=R_0+b_1 L^{y_1} \;, 
\label{fitForLogA}
\end{equation}
and in the CE with the formula 
\begin{equation}
R=R_0+\frac{b_1}{\ln L} + \frac{b_2}{(\ln L)^2} + \frac{b_3}{(\ln L)^3}  \;. 
\label{fitForLogB}
\end{equation}
The fit results are summarized in Table \ref{tab-wrapping-q2}.
The value of $R_0$ was obtained from the fit or fixed at its theoretical value 
in alternative lines, and for the CE only two correction terms were assumed
when $R_0$ was not fixed. 
It can be seen that the wrapping probabilities share the same universal values 
in the CE and the GCE, and that the leading finite-size correction term is proportional 
to $1/\ln L$ in the CE. 
Figure ~\ref{fig-R1Lq2} demonstrates the logarithmic correction by showing the data for $R_1$.
These results are consistent with the analysis Sec. \ref{sec-uc}.

\begin{table}
\caption{Fit results for the wrapping probabilities of the Ising model.
The abbreviations ``GCE" and ``CE" stand for the grand-canonical and canonical ensemble, respectively.
Entries ``-" indicate that the corresponding term is not included in the fit, 
and the numbers without error bars are fixed in the fits.
Exact values for $R_0$ were obtained using formulas in Ref.~\cite{Arguin02}.
Error margins are quoted as two times of the statistical errors in the fits.}
\label{tab-wrapping-q2}
\begin{center}
    \begin{tabular}{cllllll}
    \hline
        \multicolumn{7}{c}{GCE, Fits using $R = R_0 + b_1 L^{y_1}$} \\
      & $R_0$           & $b_1$       & $y_1$     &     & $\Lmin$ & $\chi^2/DF$ \\
    \hline
$R_x$ & $0.627\,18(15)$ & $-0.13(21)$ & $-2.1(8)$ &     & $8$     & $1.1$ \\
      & $0.627\,138\,794$     & $-0.18(26)$ & $-2.2(7)$ &     & $8$     & $0.9$ \\
$R_1$ & $0.146\,43(5)$  & $\ \; \, 0.166(20)$ & $-1.60(6)$ &     & $8$     & $0.6$ \\
      & $0.146\,436\,927$     & $\ \; \, 0.168(16)$ & $-1.60(4)$ &     & $8$     & $0.5$ \\
$R_b$ & $0.480\,75(18)$  & $-0.26(9)$ & $-1.68(16)$ &     & $8$     & $1.1$ \\
      & $0.480\,701\,867$  & $-0.27(8)$ & $-1.71(12)$ &     & $8$     & $0.9$ \\
$R_e$ & $0.773\,62(17)$  & $\ \; \, 0.09(4)$ & $-1.47(24)$ &     & $8$     & $1.3$ \\
      & $0.773\,575\,721$  & $\ \; \, 0.08(3)$ & $-1.43(16)$ &     & $8$     & $1.1$ \\
    \hline
        \multicolumn{7}{c}{CE, Fits using $R = R_0 + b_1/\ln L + b_2/(\ln L)^2 + b_3/(\ln L)^3$} \\
      & $R_0$           & $b_1$       & $b_2$       & $b_3$       & $\Lmin$ & $\chi^2/DF$ \\
    \hline
$R_x$ & $0.626\,1(12)$  & $-0.030(9)$ & $-0.013(16)$ & -  & $16$     & $0.4$ \\
      & $0.627\,138\,794$     & $-0.042(3)$ & $\ \; \, 0.029(14)$ & $-0.05(2)$  & $8$     & $0.3$ \\
$R_1$ & $0.148\,6(24)$  & $\ \; \, 0.134(24)$ & $\ \; \, 0.02(6)$ & -  & $64$     & $2.1$ \\
      & $0.146\,436\,927$  & $\ \; \, 0.162\;1(24)$ & $-0.104(18)$ & $\ \; \, 0.18(3)$  & $16$     & $1.0$ \\
$R_b$ & $0.477\,9(28)$  & $-0.166(25)$ & $-0.03(6)$ & -  & $32$     & $0.8$ \\
      & $0.480\,701\,867$  & $-0.205(6)$ & $\ \; \, 0.14(5)$ & $-0.25(8)$  & $16$     & $0.5$ \\
$R_e$ & $0.775\,6(23)$  & $\ \; \, 0.095(21)$ & $\ \; \, 0.03(5)$ & -  & $32$     & $0.1$ \\
      & $0.773\,575\,721$  & $\ \; \, 0.118(3)$ & $-0.053(14)$ & $\ \; \, 0.097(20)$  & $8$     & $0.3$ \\
    \hline
    \end{tabular}
\end{center}
\end{table}

\subsection{$2y_t-d<0$}
For $2y_t-d<0$, an example in the CE is the percolation model 
with fixed number of bonds (sites). Results for two-dimensional percolation models 
in Ref.~\cite{HBD} are consistent with our analytical results for systems $2y_t-d<0$, 
i.e. values of the universal wrapping probabilities and dimensionless ratios 
in the CE are the same with those in the GCE; and correction exponent 
$y_{\rm can}=-|2y_t-d|$ is present in the FSS. 

\subsection{The dilute Potts model} 
We also investigated the $q=3$ dilute Potts model to demonstrate 
the universality of the wrapping probabilities in the CE. 
The dilute Potts model is defined by the Hamiltonian 
\begin{equation}
{{\cal H}_d}/k_B T = -K \sum_{e_{ij}}  \delta_{\sigma_i \sigma_j} (1-\delta_{\sigma_i,0}) 
- D \sum_k \delta_{\sigma_k 0} \, . 
\end{equation}
Here $\sigma_i=0$ stands for a vacancy at site $i$, and 
$\sigma_i = 1, 2, \ldots , q$ for one of the Potts states.
The abundance of vacancies is controlled by the chemical potential $D$.
In the limit $D \rightarrow - \infty$, the vacancies are excluded and the above Hamiltonian
describes the pure Potts model. 
The partition sum of the dilute Potts model in the RC representation is 
\begin{eqnarray}
Z^{\rm d}_{\rm RC}(v,\mu, q) = 
\sum_{\calA \subseteq G} v^{\calN_{\rm b}} q^{\calN _{\rm c}} \mu ^{\calN _{\rm v}} \;. 
\label{zrc-dilute}
\end{eqnarray}
In comparison with the RC representation of the pure Potts model, a term $\mu^{\calN _{\rm v}}$ appears, 
with $\mu=e^D$, and a number $\calN _{\rm v}$ of vacancies. 
One has $\calN _{\rm v} = 0$ for the pure Potts model. 
In two dimensions, for $q<4$, the phase diagram of the dilute Potts model
in the $(v, \mu)$ plane consists of a line of Potts critical transitions and a line of first-order
transitions, which join at a tricritical point \cite{NBS79}. 
As a single fixed point governs the Potts critical transitions,
the dilute Potts model belongs to the same universality class as the pure 
Potts model. We define the dilute Potts model 
with a fixed number of Potts spins (the number of vacancies is also fixed since the total number
of sites is conserved) as the model in the CE. Our finite-size analysis
for the pure Potts model applies also to the dilute Potts model, with the bond density replaced 
by the vacancy density, and the thermal scaling field being $t \simeq A(\mu-\mu_c)$ 
($t$ should be approximated by $A_1 (\mu-\mu_c)+A_2(v-v_c)$, but here we take $v=v_c$ ). 

To simulate the dilute Potts model, in the GCE, the SW method
is used for updating Potts spins, and the Metropolis algorithm is
employed to allow fluctuations between the vacancies and Potts spins.
In the CE, the SW method is also used for updating Potts spins,
but in order to fix the total number of vacancies while still allow for
their spatial fluctuations, a Kawasaki-like algorithm is conducted, which exchange the states of
two randomly selected sites with a probability satisfying the detailed balance condition.
Replacing the SW method by the SWCM method, 
this algorithm for the CE applies generally to dilute RC models with $q>1$.
For integer $q$-state dilute Potts models in the CE,
the geometric cluster algorithm \cite{HB96} can be used, which has a more efficient
dynamics than the Kawasaki-like algorithm described above.
The simulation results at the critical vacancy density $\rho^{\rm v}_c$ in the CE were
obtained from linear interpolations between $\calN _{\rm v}=[\rho^{\rm v}_c L^2]$ and $\calN _{\rm v}=[\rho^{\rm v}_c L^2] + 1$,
where $[ \cdot ]$ denotes the integer part of the number. 

Numerical simulations were done at a critical point $K_c=1.169\;41(2)$, 
$D_c=1.376\;483\;(5)$ \cite{QDB05} for the $q=3$ dilute Potts model,
with the average vacancy density being $\rho^{\rm v}_c = 0.105\;28(1)$ \cite{QDB05}.
This point was determined such as to suppress the leading irrelevant scaling
field. Thus, in the GCE, we expect corrections with a scaling exponent smaller than $y_i=-4/5$, 
e.g. the integer $-2$ \cite{QDB05}. 
We conducted the simulations at $7$ different sizes with 
$8 \le L \le 512$. The number of samples were about $1.7 \times 10^9$ for
$L \le 64$, $2.0 \times 10^8$ for $L=128, 256$, and $6.4 \times 10^7$ for $L=512$ in the GCE;
and about $5 \times 10^7$ for each size in the CE.

\begin{figure}
\begin{center}
\includegraphics[width=8.0cm]{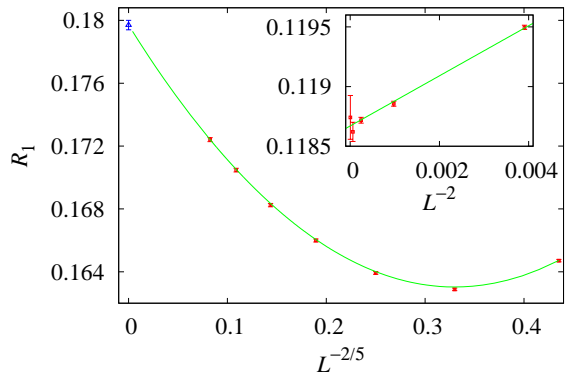}
\end{center}
\caption{Wrapping probability $R_1$ versus $L^{y_1}$ for the dilute $q=3$ Potts model 
in the canonical ensemble (main plot) and the grand-canonical ensemble (inset), 
where it is assumed that the leading correction exponent $y_1$ takes value $-2/5$ and $-2$, respectively. 
The (blue in the web version) triangular point represents the critical value of $R_1$ in the canonical ensemble 
$R^{\rm can}_{1,0} = 0.179\;7(3)$. The solid line connecting data points in the canonical ensemble
is described by $R_1(L)=0.179\;7-0.101 L^{-2/5} + 0.153 L^{-4/5}$, 
and the line in the inset is $R_1(L)=0.118\;666\;330 + 0.21 L^{-2}$.}
\label{fig-RL}
\end{figure}

We performed fits for the data of the $q=3$ dilute Potts model using ansatzes 
that are similar to those for the $q=3$ pure Potts model. 
The fit results for the wrapping probabilities are shown in Table \ref{tab-wrapping-q3-dilute}.
Similar to the case for the pure $q=3$ Potts model, the data in the GCE can also 
be well described by known exact values of the wrapping probabilities, as expected from universality; 
and the wrapping values in the CE are different from those in the GCE for the $q=3$ dilute Potts model. 
Moreover, in the CE, within uncertainties, the wrapping values for the $q=3$ pure and dilute Potts model 
agree with each other. This demonstrates that universality of the wrapping 
probability still holds in the CE. 
In contrast to the small finite-size corrections
in the GCE, the data in the CE are consistent with the appearance of correction
terms with exponents $-2/5$ and $-4/5$,
as expected from the analysis in Sec. \ref{sec-uc}. As an example, Fig.~\ref{fig-RL} illustrates 
the finite-size behavior of $R_1$.

\begin{table}
\caption{Fit results for wrapping probabilities of the $q=3$ dilute Potts model.
The abbreviations ``GCE" and ``CE" stand for the grand-canonical and canonical ensemble, respectively.
Entries ``-" indicate that the corresponding term is not included in the fit, 
and the numbers without error bars are fixed in the fits.
Exact values for $R_0$ were obtained using formulas in Ref.~\cite{Arguin02}.
Error margins are quoted as two times of the statistical errors in the fits.}
\label{tab-wrapping-q3-dilute}
\begin{center}
    \begin{tabular}{cllllllll}
    \hline
        \multicolumn{9}{c}{Fits using $R = R_0 + b_1 L^{y_1} + b_2 L^{y_2}$} \\
    \hline
& & $R_0$ & $b_1$ & $y_1$ & $b_2$ & $y_2$ & $\Lmin$ & $\chi^2/DF$ \\
    \hline
$R_x$ & GCE & $0.695\,000\,176$ & $-0.16(8)$ & $-2.1(3)$ & - & - & $8$ & $0.6$ \\
&     & $0.695\,04(6)$ & $-0.121(6)$ & $-2$ & - & - & $8$ & $0.6$ \\
& CE  & $0.667\,7(12)$ & $\ \; \, 0.045\,(15)$  & $-2/5$  & $-0.26(4)$ & $-4/5$  & $32$ & $0.5$ \\
$R_1$ & GCE & $0.118\,666\,330$ & $\ \; \, 0.29(12)$ & $-2.11(15)$ & - & - & $16$ & $0.9$ \\
&     & $0.118\,67(3)$ & $\ \; \, 0.19(4)$ & $-2$ & - & - & $16$ & $1.1$ \\
& CE  & $0.179\,7(3)$ & $-0.101\,(3)$  & $-2/5$  & $\ \; \, 0.153\,(7)$ & $-4/5$  & $16$ & $0.5$ \\
$R_b$ & GCE & $0.576\,333\,845$ & $-0.40(8)$ & $-2.08(9)$ & - & - & $8$ & $0.6$ \\
&     & $0.576\,39(7)$ & $-0.342(6)$ & $-2$ & - & - & $8$ & $0.7$ \\
& CE  & $0.487\,9(14)$ & $\ \; \, 0.15(2)$  & $-2/5$  & $-0.42(5)$ & $-4/5$  & $32$ & $0.5$ \\
$R_e$ & GCE & $0.813\,666\,506$ & $\ \; \, 0.09(4)$ & $-1.95(23)$ & - & - & $8$ & $0.6$ \\
      &     & $0.813\,68(5)$ & $\ \; \, 0.099\,(5)$ & $-2$ & - & - & $8$ & $0.6$ \\
& CE  & $0.847\,6(10)$ & $-0.058(13)$  & $-2/5$  & $-0.1(4)$ & $-4/5$  & $32$ & $0.4$ \\
    \hline
    \end{tabular}
\end{center}
\end{table}

\subsection{Dimensionless ratios}
We also observed two dimensionless ratios $Q_\calC$, $Q_\calS$ which are defined as following:
\begin{equation}
Q_\calC= \frac{ \langle \calC_1 \rangle^2}{\langle  \calC_1^2 \rangle} \; ,
\end{equation}
where $\calC_1$ is the size of the largest cluster;
\begin{equation}
Q_\calS=\frac{(q-1) \langle \calS_2 \rangle^2}{\langle (q+1) \calS_2^2-2\calS_4 \rangle},
\end{equation}
where $\calS_k= \sum_{j=1}^{\calN_{\rm c}} s_j^k$, with $s_j$ being the size of
the $j$th cluster divided by the volume $L^2$.
For $q=2$, the latter quantity is equivalent to the Binder ratio \cite{Binder81}. 
It is found that these ratios behave similar to the wrapping probability. 
For both the $q=3$ pure and dilute Potts model, which have $2y_t-d > 0$, 
the critical values of the ratios in the CE are different from those in the GCE, 
but they remain universal. Table~\ref{tab-RQ} summarizes the critical values of
the two dimensionless ratios and the wrapping probabilities for the $q=3$ Potts models. 

\begin{table}
\caption{Critical value of the wrapping probabilities and dimensionless ratios
in the thermodynamic limit $L \rightarrow \infty$, for the $q=3$ pure and dilute Potts model
in the grand canonical ensemble (GCE) and the canonical ensemble (CE).
These results tell that the CE values are different from the GCE values,
and the differences are universal. 
The exact values were obtained using formulas in Ref.~\cite{Arguin02}.}
\label{tab-RQ}
\begin{center}
    \begin{tabular}{llll}
    \hline
    model      & $R_x$ & $R_1$ & $R_b$  \\
    \hline
    GCE pure & $0.695\;2 (3)$ & $0.118\;68 (10)$ & $0.576\;6 (2)$ \\
    GCE dilute & $0.695\;04 (6)$ & $0.118\;67 (3)$ & $0.576\;39 (7)$ \\
    GCE exact    &  $0.695\;000\;176$  & $0.118\;666\;330$    & $0.576\;333\;845$   \\
    CE pure & $0.667\;3 (5)$ & $0.180\;1 (2)$ & $0.487\;1 (7)$   \\
    CE dilute & $0.667\;7 (12)$ & $0.179\;7 (3)$ & $0.487\;9 (14)$ \\
    \hline
    model      & $R_e$ & $Q_\calC$ & $Q_\calS$   \\
    \hline
    GCE pure & $0.813\;8 (2)$ & $0.936\;36 (14)$ &  $0.854\;3(3)$ \\
    GCE dilute  & $0.813\;68 (5)$ & $0.936\;26 (14)$ &  $0.854\;1(3)$ \\
    GCE exact    &  $0.813\;666\;506$   &                & \\
    CE pure & $0.847\;4 (3)$ & $0.974\;2 (2)$ &  $0.919\;2 (5)$ \\
    CE dilute  & $0.847\;6 (10)$ & $0.974\;3 (4)$ &  $0.919\;5 (9)$ \\
    \hline
    \end{tabular}
\end{center}
\end{table}

\section{Summary and discussion} 
\label{sec-sum}

In this work, based on FSS analysis, 
we derive that for systems with $2y_t-d>0$ in the GCE, in the limit of $L \rightarrow \infty$,
the universal wrapping probabilities at criticality become different in the CE.  
However, they are still universal in the CE. 
For $2y_t-d \le 0$, the critical universal wrapping probabilities do not change. 
It is also derived that, for $2y_t-d \ne 0$, finite-size corrections with exponent 
$y_{\rm can} = -|2y_t-d|$ appear in the CE. 
For $2y_t-d=0$, i.e. the specific heat has a logarithmic anomaly in the GCE, 
the leading correction term changes to a logarithmic form in the CE, 
such as $1/\ln L$ for the two-dimensional Ising model. 
Other dimensionless quantities, such as the Binder ratio, behave similar to
the wrapping probabilities. These results are supported by numerical results from 
extensive Monte Carlo simulations. 

The simulations were conducted for two-dimensional critical systems,
including the $q=3$ Potts model ($y_t=6/5$), the Ising model ($y_t=1$),
and the percolation model ($y_t=1/\nu=3/4$).
As described in Sec. \ref{sec-mod}, these models can be described in the RC representation of
the $q$-state Potts model, which are well defined also for positive noninteger $q$.
In two dimensions, exact critical exponents for the Potts model can
be obtained from the Coulomb gas theory \cite{Nienhuis87} 
or the conformal invariance theory \cite{Cardy87}.
In the former case, they can be expressed in the Coulomb gas coupling constant $g$ 
which depends on $q$ by
\begin{equation} 
q = 2 + 2 \cos \frac{g \pi}{2} \; ,
\end{equation}
with $2 \le g \le 4$ for the critical Potts model.
The thermal renormalization exponent $y_t$ is given by
\begin{equation}
X_t = d-y_t = \frac{6}{g} -1 \;. 
\end{equation}
Thus, for the critical Potts model in two dimensions, 
one has  $2y_t-d>0$ for $2 < q \le 4$, $2y_t-d=0$ for $q=2$, and $2y_t-d<0$ for $0 \le q < 2$.
In higher dimensions, the Ising model has $2y_t-d \simeq 0.174$ for $d=3$ \cite{Hasenbusch10}, 
and the percolation model has $2y_t-d <0$ also for $d > 2$ \cite{WZZGD, BLJ}. 
The results summarized in the previous paragraph should apply but not restricted to these models,
as long as the wrapping probability can be properly defined (wrapping is trivial on a complete graph).
It should be mentioned that, for the percolation model, since the particle density has no finite-size dependence, 
the derivation presented need some modifications \cite{HBD}, which do not change the results 
for the wrapping probability. 

Exact values for the critical universal wrapping probability of the Potts model in the GCE 
were obtained through the analysis of the homology group of the torus based on a method 
introduced by di Francesco et al.\ \cite{FSZ87, Pinson94, Arguin02, MS09, TB14} 
(Ref. \cite{TB14} contains a brief review).
It remains an open question whether exact results can be obtained for systems with $2y_t-d>0$ in the CE. 
Although critical universal wrapping probabilities and dimensionless ratios remain 
universal in the CE, some universal parameters may become nonuniversal,
such as the excess cluster number for percolation \cite{HBD}, 
which is universal in the GCE. It would be interesting to study properties of other quantities, 
and constraints other than the constraint on the number of particles, e.g. fixing the magnetization. 

Finally, we note that the wrapping probabilities and dimensionless ratios considered in this work 
are different from the universal quantities studied by Izmailian and Kenna \cite{IK}, which are expressed as
combinations of universal amplitudes governing the behavior of various quantities near the critical point. 

\section*{Acknowledgments}
We thank H. W. J. Bl\"ote for the collaboration on Ref. \cite{HBD} 
and a paper \cite{DB05} which initiated this work, and for his critical reading of the manuscript. 
We also thank R. Ziff for his comments.
This work is supported by the National Natural Science Foundation of China under Grant No. 11275185, 
and by the Open Project Program of State Key Laboratory of Theoretical Physics, Institute of Theoretical Physics,
Chinese Academy of Sciences, China (No. Y5KF191CJ1). 
Y. Deng acknowledges the Ministry of Education (China) for 
the Specialized Research Fund for the Doctoral Program of Higher Education under Grant No. 20113402110040 
and the Fundamental Research Funds for the Central Universities under Grant No. 2340000034. 

\section*{References}
\bibliographystyle{elsarticle-num}

\begin{thebibliography}{99}

\bibitem{LPPA}
R. P. Langlands, C. Pichet, Ph. Pouliot, and Y. Saint-Aubin,
J. Stat. Phys. {\bf 67} (1992) 553.
\bibitem{Pinson94}
H. T. Pinson, J. Stat. Phys. {\bf 75} (1994) 1167.
\bibitem{ZLK99}
R. M. Ziff, C. D. Lorenz, and P. Kleban, Physica (Amsterdam) {\bf 266A} (1999) 17.
\bibitem{NZ}
M. E. J. Newman and R. M. Ziff, Phys. Rev. Lett. {\bf 85} (2000) 4104;
Phys. Rev. E {\bf 64} (2001) 016706.
\bibitem{WZZGD}
J. Wang, Z. Zhou, W. Zhang, T. M. Garoni, Y. Deng, Phys. Rev. E {\bf 87} (2013) 052107.
\bibitem{FDB}
X. Feng, Y. Deng, and H. W. J. Bl\"ote, Phys. Rev. E {\bf 78} (2008) 031136.   
\bibitem{HBD}
H. Hu, H. W. J. Bl\"ote, and Y. Deng, J. Phys. A: Math. Theor. {\bf 45} (2002) 494006.
\bibitem{Arguin02}
L. P. Arguin, J. Stat. Phys. {\bf 109} (2002) 301.
\bibitem{MS09}
A. Morin-Duchesne and Y. Saint-Aubin, Phys. Rev. E. {\bf 80} (2009) 021130.
\bibitem{TB14}
T. Blanchard, J. Phys. A: Math. Theor {\bf 47} (2014) 342002.
\bibitem{LDGB}
Q. Liu, Y. Deng, T. M. Garoni, and H. W. J. Bl\"ote, Nucl. Phys. B {\bf 859} (2012) 107.
\bibitem{PC87}
E. L. Pollock and D. M. Ceperley, Phys. Rev. B {\bf 36} (1987) 8343.
\bibitem{PHA} 
V. Privman, P. C. Hohenberg, and A. Aharony, in {\it Phase Transitions 
and Critical Phenomena}, edited by C. Domb and J. L. Lebowitz, 
(Academic Press, London, 1991), Vol. 14, p. 1.
\bibitem{PV02}
A. Pelissetto and E. Vicari, Physics Reports {\bf 368} (2002) 549.
\bibitem{Binder81}
K. Binder, Z. Phys. B - Condensed Matter {\bf 43} (1981) 119.
\bibitem{BinderStu}
G. Kamieniarz and H. W. J. Bl\"ote, J. Phys. A: Math. Gen. {\bf 26} (1993) 201; \\
Y. Okabe, K. Kaneda, M. Kikuchi, and C.-K. Hu, Phys. Rev. E. {\bf 59} (1999) 1585; \\
W. Selke and L. N. Shchur, Phys. Rev. E {\bf 80} (2009) 042104; \\
B. Kastening, Phys. Rev. E {\bf 87} (2013) 044101; \\
A. Malakis, N. G. Fytas, and G. G\"ulpinar, Phys. Rev. E {\bf 89} (2014) 042103. 
\bibitem{Wu82}
F. Y. Wu, Rev. Mod. Phys. {\bf 54} (1982) 1.
\bibitem{Fisher68}
M. E. Fisher, Phys. Rev. {\bf 176} (1968) 257.
\bibitem{IK}
N. Sh. Izmailian and R. Kenna, J. Stat. Mech. (2014) P07011;
Condensed Matter Phys. {\bf 17} (2014) 33602.
\bibitem{KHF}
R. Kenna, H. P. Hsu, and C. von Ferber, J. Stat. Mech. (2008) L10002.
\bibitem{DengThesis}
Y. Deng, {\it Conformal Symmetries and Constrained Critical Phenomena} (Delft University Press,
Delft, 2004).
\bibitem{Krech}
M. Krech, in {\it Computer Simulation Studies in Condensed Matter Physics XII}, ed. D. P. Landau,
S. P. Lewis, and H. B. Schuettler (Springer, Berlin, 1999), arXiv:cond-mat/9903288.
\bibitem{MOF}
I. M. Mryglod, I. P. Omelyan, and R. Folk, Phys. Rev. Lett. {\bf 86} (2001) 3156.
\bibitem{MrFo}
I. M. Mryglod, R. Folk, Physica A {\bf 294} (2001) 351.
\bibitem{KF}
P. W. Kasteleyn and C. M. Fortuin J. Phys. Soc. Jpn. {\bf 26} (Suppl.) (1969) 11;
C. M. Fortuin and P. W. Kasteleyn Physica (Amsterdam) {\bf 57} (1972) 536.
\bibitem{PF84}
V. Privman and M. E. Fisher, Phys. Rev. B {\bf 30} (1984) 322.
\bibitem{Onsager44}
L. Onsager, Phys. Rev. {\bf 65} (1944) 117.
\bibitem{SW}
R. H. Swendsen and J.-S. Wang, Phys. Rev. Lett. {\bf 58} (1987) 86.
\bibitem{Kawasaki}
K. Kawasaki, in {\it Phase Transitions and Critical Phenomena} Vol. 2,
edited by Domb C and Green M (Academic, London, 1972)  p. 443.
\bibitem{CM}
L. Chayes and J. Machta, Physica (Amsterdam) {\bf 254A} (1998) 477.
\bibitem{Nienhuis87}
B. Nienhuis, in {\it Phase Transitions and Critical Phenomena}, edited by C. Domb and J. L. Lebowitz, 
(Academic Press, London, 1987), Vol. 11, p. 1.
\bibitem{Cardy87}
J. L. Cardy, in {\it Phase Transitions and Critical Phenomena}, edited by C. Domb and J. L. Lebowitz, 
(Academic Press, London, 1987), Vol. 11, p. 55.
\bibitem{NBS79}
B. Nienhuis, A. N. Berker, E. K. Riedel, and M. Schick,
Phys. Rev. Lett. {\bf 43} (1979) 737.
\bibitem{HB96}
J. R. Heringa and H. W. J. Bl\"ote, Physica A {\bf 232} (1996) 369;
Phys. Rev. E {\bf 57} (1998) 4796.
\bibitem{QDB05}
X. Qian, Y. Deng, and H. W. J. Bl\"ote, Phys. Rev. E {\bf 72} (2005) 056132.
\bibitem{Hasenbusch10}
M. Hasenbusch, Phys. Rev. B {\bf 83} (2010) 174433 and reference therein.
\bibitem{BLJ}
H. G. Ballesteros, L. A. Fernandez, V. Martin-Mayor, A. Munoz Sudupe, G. Parisi,
and J. J. Ruiz-Lorenzo, Phys. Lett. B {\bf 400} (1997) 346; \\
C. D. Lorenz and R. M. Ziff, Phys. Rev. E {\bf 57} (1998) 230; \\
N. Jan, D. C. Hong, and H. E. Stanley, J. Phys. A {\bf 18} (1985) L935.
\bibitem{FSZ87}
P. di Francesco, H. Saleur, J. B. Zuber, J. Stat. Phys. {\bf 49} (1987) 57-79.
\bibitem{DB05}
Y. Deng, and H. W. J. Bl\"ote, arXiv:cond-mat/0508348.

\end{thebibliography}

\end{document}